\documentclass[manuscript]{aastex}

\usepackage{xspace}

\newcommand{\mysource}{IRAS 18113$-$2503\xspace}
\newcommand{\kms}{km s$^{-1}$\xspace}
\newcommand{\vlsr}{$V_{\rm LSR}$\xspace}

\shorttitle{water fountain IRAS 18113-2503}
\shortauthors{Jos\'e F. G\'omez et al.}

\begin{document}

\title{IRAS 18113-2503: the water fountain with the fastest jet?}

\author{Jos\'e F. G\'omez\altaffilmark{1,2}, J. Ricardo Rizzo\altaffilmark{3},
Olga Su\'arez\altaffilmark{4}, Luis F. Miranda\altaffilmark{5,6},
Mart\'{\i}n A. Guerrero\altaffilmark{1}, Gerardo Ramos-Larios\altaffilmark{7}}

\altaffiltext{1}{Instituto de Astrof\'{\i}sica de Andaluc\'{\i}a,
  CSIC, Apartado 3004, E-18080 Granada, Spain; e-mail: jfg@iaa.es}
\altaffiltext{2}{On sabbatical leave at the Australia Telescope National 
Facility, CSIRO Astronomy \& Space Science, Marsfield, NSW 2122, Australia }
\altaffiltext{3}{Centro de Astrobiolog\'{\i}a, CSIC-INTA, E-28850 Madrid,
Spain} 
\altaffiltext{4}{UMR 6525 H. Fizeau, Universit\'e de Nice Sophia
  Antipolis, CNRS, OCA. Parc Valrose, 06108 Nice Cedex 2, France}
\altaffiltext{5}{Consejo Superior de Investigaciones Cient\'{\i}ficas, 
C/ Serrano 117,
E-28006 Madrid, Spain}
\altaffiltext{6}{Departamento de Fisica Aplicada, Facultade de Ciencias,
Universidade de Vigo, E-36310 Vigo, Spain (present address)}
\altaffiltext{7}{Instituto de Astronom\'{\i}a y Meteorolog\'{\i}a, 
Av. Vallarta No. 2602, Col. Arcos Vallarta, C.P. 44130 Guadalajara, 
Jalisco, Mexico}

\begin{abstract}
We present Expanded Very Large Array (EVLA) water maser observations at 22 GHz toward the source
\mysource. Maser
components span over a very high velocity range of $\simeq 500$
\kms, the second largest found in a Galactic maser, only
surpassed by the high-mass star forming region W49N. 
Maser components are grouped into a blue and a redshifted cluster, 
separated by $0.12''$. Further mid-IR and radio data suggest that
\mysource is a post-AGB star, thus a new bona fide member of the rare
class of ``water fountains''. It is the evolved object with the
largest total velocity spread in its water masers, and with the
highest velocity dispersion within its red- and blue-shifted
lobes ($\simeq 170$ \kms). The large total velocity range of 
emission probably indicates that IRAS\,18113-2503 
has the fastest jet among the known water fountain stars. On the other hand, 
the remarkably high velocity dispersion within each lobe 
may be interpreted in terms of shocks produced by an episode of mass ejection 
whose velocity increased up to very high values or, alternatively, by
projection effects in a jet with a large opening angle and/or
precessing motions.

\end{abstract}

\keywords{masers; stars: AGB and post-AGB; stars: individual
  (\mysource ); stars: mass-loss; stars: winds, outflows}

\section{Introduction}

Water fountain (WF) stars  are evolved objects with water maser emission
spreading over large velocity ranges, $\ga 100$ \kms \citep[][for a
review]{lik88,ima07}.  Interferometric observations of water
maser components in WFs 
always show a bipolar
distribution, with well defined, spatially separated clusters of blue-
and redshifted features \citep[e.g.,][]{ima04,bob05}. Moreover, proper motion measurements of
these water masers indicate that they trace very collimated jets, with
extremely short timescales $\la 100$ years \citep[see, e.g.,][]{ima04}. WFs seem to be in the
transition between the Asymptotic Giant Branch (AGB) and the
planetary nebula (PN) stage, in the phase known as post-AGB, although 
W43A might still be in the AGB \citep{ima02}.

In the late evolution of low- and intermediate-mass stars ($<8-10$ M$_\odot$),
the morphology of the detached shells in AGB stars are mostly spherical \citep{olo10}, 
while many PNe show clear departures from sphericity, with bipolar or
multipolar morphologies. It has been proposed that highly collimated outflows, 
generated in the late AGB and
post-AGB phases could be the main shaping mechanism of PNe
\citep{sah98}. 
In this context, WFs could be key objects to
understand these shaping processes, since their jets represent one of the 
 first manifestations of collimated
mass-loss in evolved low- and intermediate-mass stars. 

The number of known WFs is scarce, given the very short timescale 
of their jets, and 
that these objects still retain a thick circumstellar envelope,
which renders most of them invisible at optical wavelengths. Only
13 candidate WFs have been reported so far,  nine of which have been
confirmed with interferometric water maser observations
(IRAS 16342$-$3814, \citealt{cla09}; IRAS 16552$-$3050,
\citealt{sua08}; IRAS 18286$-$0959, IRAS 18460$-$0151,
\citealt{ima07}; 
IRAS 19134+2131, \citealt{ima04}; IRAS
19190+1102, \citealt{day10}; OH
009.1-0.4, \citealt{wal09}; OH 12.8$-$0.9, \citealt{bob05}; W43A,
\citealt{ima02}). We note that interferometric observations of 
masers
are crucial to properly classify a source as a WF. Single-dish observations may not have enough positional
accuracy to determine whether the maser emission actually arises from
an evolved object or from a nearby source (for instance, a
star-forming region), 
or whether the observed spectrum is the
superposition of the emission from several sources.

During a single-dish survey for water maser emission towards
optically obscured candidate post-AGB stars and PNe (G\'omez et al.,
in preparation), we discovered a striking spectrum toward the position
of \object{\mysource},
whose maser components span a velocity range $\simeq 500$ \kms that,
if confirmed, would be one of the largest found in Galactic
masers. Obviously, this source merited further investigation, since it
could represent a singular case of extremely high velocity WF.

There is very limited information about this source in the
literature. It was cataloged by \cite{pre88} as a possible PN (object
PM 1-221), based on its IRAS data.
It was later listed as a candidate proto-PN 
\citep{hu93,koh01,szc07}. Its possible nature as a proto-PN (i.e, a
post-AGB star) certainly suggests that this could be a new bona fide
water fountain.

In this paper, we present new interferometric observations of water masers, carried out with the
Expanded Very Large Array (EVLA), to confirm the
nature of \mysource 
as a water fountain star, and to determine the spatio-kinematical
distribution of its maser emission. These observations have allowed us to study the particular properties
of this source with extremely high-velocity emission.

\section{Observations}

Observations of the $6_{16}-5_{23}$  transition of the water molecule (rest
frequency = 22235.08 MHz) were carried with the EVLA \citep[see
][]{per11} of
the National Radio Astronomy Observatory\footnote{The National Radio
  Astronomy Observatory is a 
  facility of the National Science Foundation operated under
  cooperative agreement by Associated Universities, Inc.} in its D configuration
on 2010 May 29, towards \mysource (EVLA project 10A-263). 
The phase center of the
observations was located at R.A.(J2000) $= 18^h 14^m 27.26^s$,
Dec(J2000) $=-25\arcdeg 03' 00.4''$, the position of the near infrared
counterpart proposed by \citet{ram09} for \mysource. 
In order to obtain sufficient
velocity coverage and spectral resolution for
this source, we set the system to simultaneously observe two
overlapping bands in dual polarization, of 320 MHz width each, 
centered at velocities with
respect to the local standard of rest (\vlsr) of 40 and 140
\kms, respectively. The resulting combined \vlsr coverage was from
$-174.8$ to 356.5
\kms. Each band was sampled over 256 channels of 125
kHz width (1.7 \kms ). We used the
source J1331+305 as flux calibrator (assumed flux density = 2.57 Jy, using
the VLA 1999.2 coefficients). 
The source J1820$-$2528 was the phase calibrator
(bootstrapped flux density = $0.92\pm 0.05$ Jy), while J1924$-$2914 was used as bandpass
calibrator (bootstrapped flux density = $13.3\pm 0.4$ Jy).

All calibration and
data reduction was carried out using the Astronomical Image Processing
System (AIPS). We followed the current recommendations for reduction
of both high-frequency and EVLA data shown in the AIPS cookbook. Data
were self-calibrated using the channel with maximum emission in each
band, and then applying the phase and amplitude solution to all
channels. Maps were obtained with task IMAGR, using a robust weighting
scheme of visibilities (with ROBUST = 0) and the CLEAN algorithm for
deconvolution.  Hanning spectral smoothing was applied to mitigate
ringing in the bandpass due to the Gibbs effect in the presence of
strong, narrow maser lines. The final effective spectral resolution
after smoothing is $\simeq 3.4$ \kms.
The resulting full width at half maximum (FWHM) of the
synthesized beam was $5''\times 2''$, P.A. = $-16\arcdeg$, and the
noise level in each channel was $\simeq 0.7$ mJy beam$^{-1}$.
The position of the maser emission in each channel was
derived by fitting elliptical Gaussians with task JMFIT. 

Following
this procedure, we found a significant difference in the nominal
position of the emission of
the same maser components in the overlapping range of both observed
bands. This difference is mainly due to atmospheric phase
fluctuations. We used the emission at \vlsr $\simeq 201$ \kms, which is
present in both bands, to align the whole data set to a common
reference position. To make the emission at this velocity coincide in
both bands, we shifted the position of all channels in the band centered at \vlsr
$=40$ \kms by $1.13''$. We 
consider this as a rough estimate of the absolute positional
error in our observations. After the realignment, the relative
positional ($1\sigma$) error among different maser components in the
map is given by $\sigma_{\rm pos} = \theta/(2 \times {\rm snr})$, where $\theta$ is
the FWHM of the synthesized beam, and snr is the signal-to-noise ratio
of the maser components. Since most of the maser components found had
flux densities $> 1$ Jy, the relative positional accuracy was
typically $\sigma_{\rm pos} < 4$ mas.

In order to further study the nature of this source, we have complemented 
the EVLA data with public archive data from the Wide-field Infrared
Survey Explorer (WISE), The Midcourse Space Explorer (MSX), AKARI, and
the Infrared Astronomical
Satellite (IRAS) space missions.

\section{Results and discussion}

\subsection{Spatial distribution and kinematics of water masers}

\begin{figure}
\epsscale{1}
\plotone{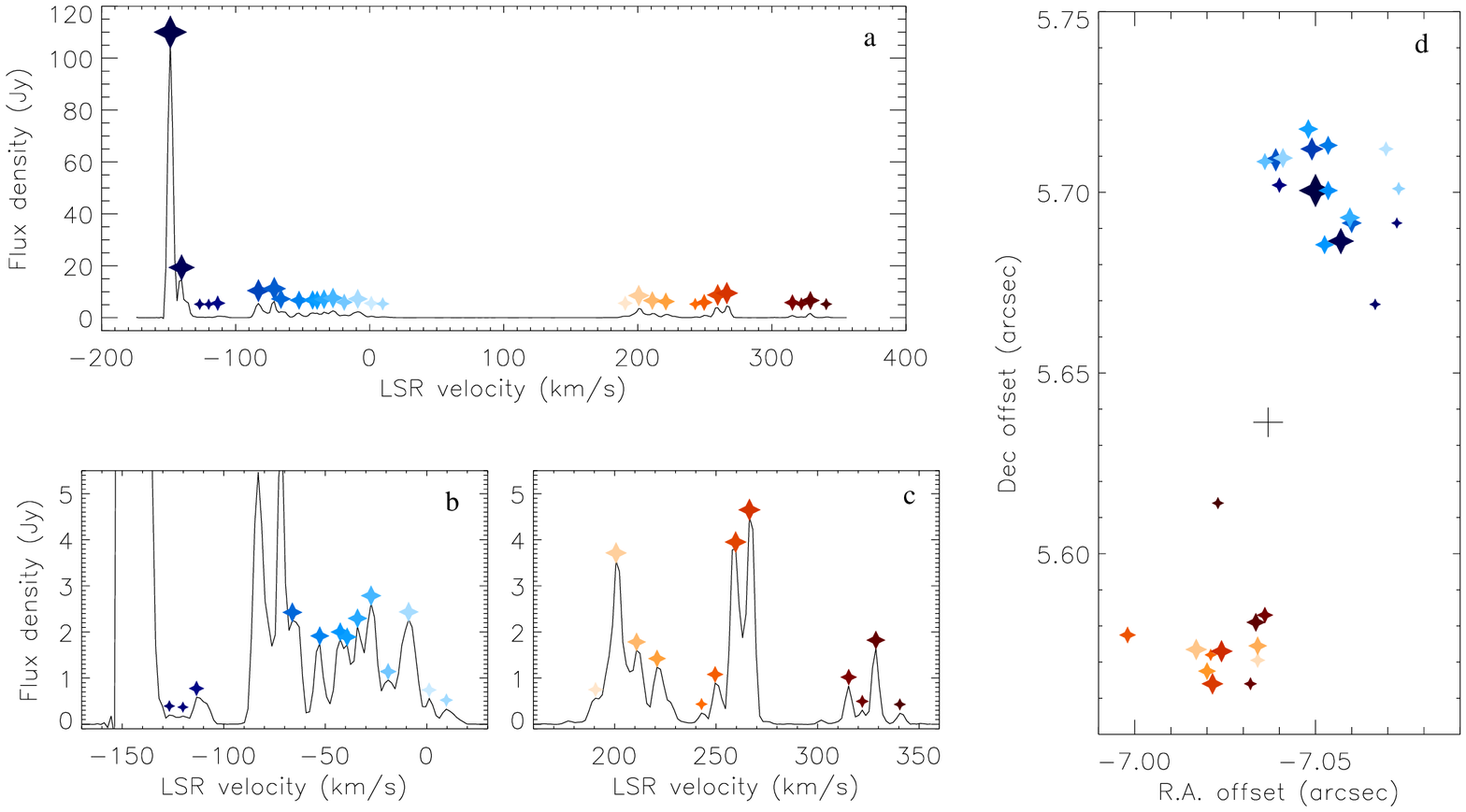}
\caption{(a) Spectrum of the water maser towards \mysource,
  integrated over an area of 218.5 arcsec$^2$. Colored stars are used
  to identify 
each feature in the rest of the panels. (b) Close-up
  of the blueshifted components. (c) Close-up of the redshifted
  components. 
(d) Map of water maser features $> 0.1$ mJy in \mysource. Axes represent
  spacial offsets with respect to the phase center of the
  observations (R.A.(J2000) $= 18^h 14^m 27.26^s$,
Dec(J2000) $=-25\arcdeg 03' 00.4''$). The size of the stars is proportional to the logarithm of the flux density of
 the components. 
 The black cross marks the estimated position of the 
central source.\label{spectrum}\label{map}}
\end{figure}

We detected water maser emission in a range from \vlsr $= -148.7$ to 350.6
\kms. The integrated spectrum in the whole field is shown in Fig.\
\ref{spectrum}a. 
The position of all maser components is within $0.15''$
(Fig. \ref{map}d), and their spatial
distribution is clearly bipolar, clustered in two
clumps, with the
blueshifted components ($-148.7$ to 9.7 \kms) to the north and the redshifted
ones (177.1 to 350.6 \kms) to the south. The intensity-weighted centroids of the blue- and redshifted
clusters are separated by $0.12''$, with P.A. $=-14\arcdeg$, a separation similar to
that found in other water fountain sources \citep{ima07}. This clearly suggests
that all components arise from the same source, whose position and radial
velocity should be approximately midway between those of the two maser
clusters. This would mean that the powering source of the maser emission
should be  at R.A.(J2000) $= 18^h 14^m 26.74^s$,
Dec(J2000) $=-25\arcdeg 02' 54.8''$ (absolute error $\simeq
1.2''$), with \vlsr $\simeq 100$ \kms. This source would be powering a
collimated jet, traced by the maser emission, with a jet velocity of
at least $\simeq 250$ \kms (half the full velocity range). Neither
maser cluster shows a well-defined morphology, like the arcs seen in
other water fountain sources \citep[e.g.,][]{bob07,day10}, suggestive
of bow shocks. They do not show an evident velocity pattern either, although
we found a weak correlation in the northern clump, where maser
components closer to the central source tend to trace higher velocities (the
value of the Spearman's rank correlation coefficient of velocity
vs. distance is $-0.52$, $p=0.03$).

Since all components seem to arise from the same source, we
note that the total velocity range of the emission ($\simeq 500$ \kms)
is the second largest ever found in a maser source in the Milky Way, 
short of the $\simeq 530$ \kms detected in the high-mass star-forming region
W49N \citep{mor76,wal82}. \mysource is the evolved star (see section
\ref{nature}) showing the largest velocity range in their maser emission,
superseding by far OH 009.1$-$0.4 \citep[$\simeq 400$ \kms,][]{wal09}. 
It is possible that there are maser
features at even higher velocities. However, our single-dish observations
with the Green Bank Telescope, taken on 2010 February
(G\'omez et al. in preparation), covered a wider range of velocities
($-237$ to 557 \kms), but detected maser components in the
same velocity range reported in this 
paper. Therefore, higher velocity components, if present, should be
significantly weaker, below the sensitivity of Green Bank.

\mysource is also remarkable in the large velocity dispersion within
each lobe of the maser jet (see Fig. \ref{spectrum}b,c), 
the largest ever found in a water fountain
source. 
There are several possible explanations for this unusually high
velocity dispersion. 
A possibility is that the jet has large opening angle, so that gas moving on trajectories
closer to the plane of the sky with have lower
relative radial velocities.
We note that the angular size of the
cluster indicate an apparent opening angle of the jet
 of $\simeq 35^\circ$, which is
relatively large compared with jets in other water fountains, although
it is an upper limit to the real opening angle. A conical jet with  
$35^\circ$ moving
at a constant velocity of $\simeq 300$ \kms could have dispersions $>
150$ \kms due to projection effects only. A further increase in
velocity dispersion could be produced if the jet is precessing (or
wobbling). Corkscrew-like
motions in masers have been observed in other water fountains \citep{ima07}.

An alternative explanation is that the large velocity dispersion in
the maser emission is an
intrinsic characteristic of the jet.
This dispersion is reminiscent of the large velocity
width found in the optical knots of the proto-PN Hen 3-1475
\citep{rie03}, $400-1000$ \kms. \citet{rie03} interpreted
the unusual spectral characteristics of these knots  as the result of 
internal shocks along the jet, generated by 
episodes of mass loss whose velocity increased with
time \citep{rag90}. In this scenario, faster material  will reach the
slower one previously ejected, thus producing a shock that can have  large
velocity dispersions.
We speculate  that the water masers in \mysource could be the result of an
ejection event in which the velocity increased up to a very high
velocity, leading to the large velocity dispersion observed in each
lobe. Moreover, the  presence of recurrent ejection events could lead to
shocks and eventually, to maser emission in different locations along the
jet (not only at the working surface of its head). This could be
the case in other
water fountain sources \citep{cla09,day10}. Whether recurrent 
ejections are common in these objects could be elucidated with 
time-monitoring studies of their maser emission.

\subsection{The nature of the central star}
\label{nature}

Water maser emission can be present in both, young and evolved
stars \citep{eli92}. A usual problem in water-maser-emitting objects is to
ascertain its true evolutionary stage. In the case of \mysource, the
high-velocity maser emission could be compatible with arising from a
high-mass young stellar object (YSO). If this is the case, we would
expect it to be associated with a molecular cloud. However, CO(J$=2-1$)
observations toward this source (Rizzo et al. in preparation) failed
to detect significant emission, with a stringent upper limit to its main beam
brightness temperature, $3\sigma = 30$ mK. 
Therefore, we can rule out the YSO hypothesis, and consider
the non-detection of CO as a strong proof that \mysource is a true water
fountain evolved star. 

\begin{figure}
\epsscale{1}
\plotone{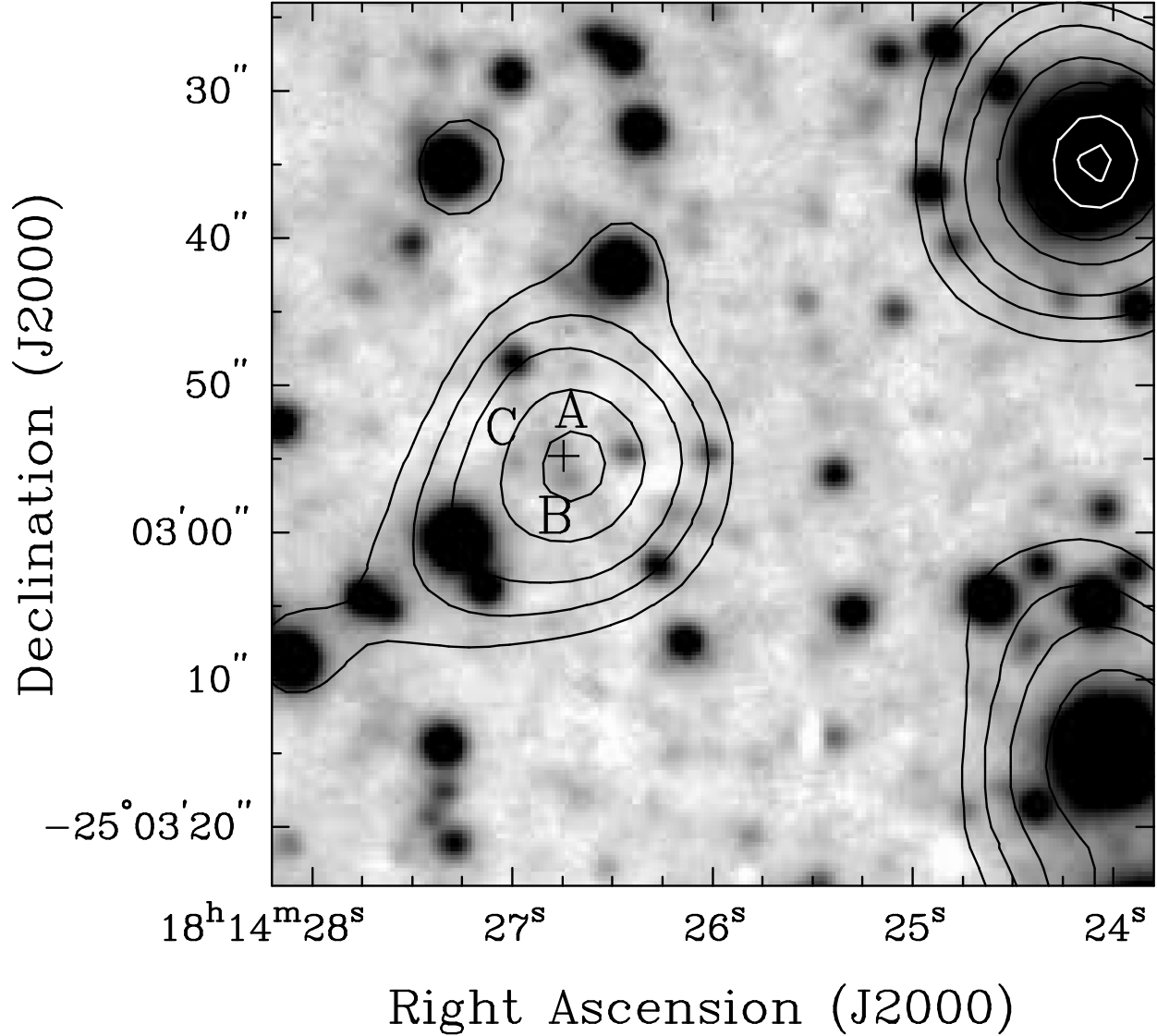}
\caption{Contour map of emission at 4.6 $\mu$m from {\em WISE}, superposed on a
  near-IR image at 2.2 $\mu$m \citep{ram11}. The cross marks the
  estimated position of 
  the source pumping the maser emission. The 2.2 $\mu$m sources
  closest to this position are labeled with letters A, B, and C. \label{msx}}
\end{figure}

Infrared data can shed some further light on the nature of the central star.
There are three weak
near-IR objects close to this source
(labeled A to C in Fig. \ref{msx}).  Neither near-IR \citep{ram11}, 
nor Johnson R and I images we took on 2010 July 28 with CAFOS at the 2.2\,m
telescope of the Calar Alto Observatory, show that any of these
objects
have peculiar colors that could suggest them to be
the powering source of the maser emission. However, at longer
wavelengths, WISE data between
3.4 and 22 $\mu$m show a
source with red colors, WISEP J181426.70-250255.6, whose position is
compatible with the 
maser emission (Fig. \ref{msx}), and with a mid-IR source
detected by MSX (source MSX6C G006.5661-03.6398) and AKARI (source
AKARI-IRC-V1 J1814267-250256 in the Point Source Catalog, and
AKARI-FIS-V1 J1814268-250304 in the Bright Source Catalog).

\begin{figure}
\includegraphics[angle=-90,scale=0.6]{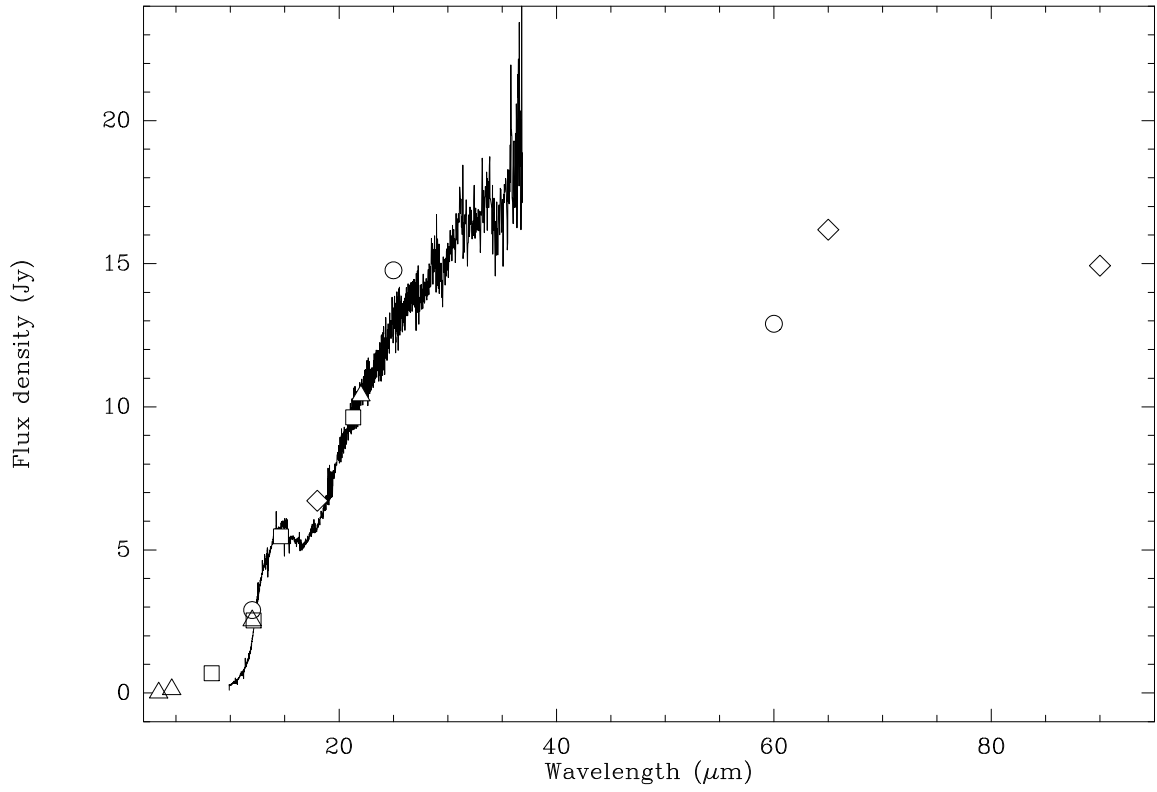}
\caption{Infrared spectra and photometry for \mysource. The solid
  black line is the Spitzer spectrum. Squares, circles, 
 diamonds, and triangles are MSX, IRAS, AKARI, and WISE data, 
respectively. \label{sed}}
\end{figure}

A mid-IR spectrum of 
MSX6C G006.5661-03.6398, extracted from
the Spitzer Heritage archive is shown in Fig \ref{sed}, together with
all photometric points found in the MSX, AKARI, IRAS, and WISE
catalogs. The Spitzer spectrum shows a broad amorphous silicate
absorption at 18 $\mu$m, typical
of O-rich stars. Moreover, we note that the photometric data have flux
density values very close to those in the Spitzer spectrum. Since
those infrared data were taken at different epochs, the flux stability
of the source indicates that it is not variable, thus suggesting a post-AGB
nature for the central star. Therefore, this further confirms that
\mysource is a new bona fide water fountain. 

\subsection{Comparison with other water fountain sources}

\mysource stands out from other water fountains \citep{ima07} 
as being the one with
the largest spread of line-of-sight velocities in their maser
emission (500 \kms), $\simeq 100$ and 200 \kms larger than the first and second 
runner-ups, OH 009.1$-$0.4 \citep{wal09} and IRAS
18460$-$0151 \citep{deg07}, respectively. 
Large maser velocities may indicate that they trace a jet oriented
almost along the line of sight. This could certainly be the case of
OH 009.1$-$0.4, for which the bipolar morphology detected by \citet{wal09} 
is not as clear
as that seen in
other water fountains. However, \mysource does show a distinctive
bipolar pattern, without any overlapping between the red- and
blueshifted lobe. 

A study of the proper motions of the maser spots would provide
conclusive estimates of  
the spatial velocity of the jet. 
Proper motions have been measured in other
water fountains with Very Long Baseline Interferometry
\citep[e.g.,][]{bob07,ima07,cla09,day10}. These 
observations also allow the precise determination of distances via
parallax measurements, and therefore, the full 3D velocity can be directly
measured with high accuracy. 
The fastest water fountain jet reported with this technique is that
of IRAS 18460$-$0151 \citep{ima07}, with a velocity of $\simeq 300$
\kms. However, we note that this estimate may be affected by large
uncertainties, 
since large proper motions were measured only for one side of the jet.
In the case of \mysource, the projection of the jet velocity on the line
of sight can be assumed to be at least half of the total velocity span of the maser, i.e., 
$\simeq 250$ \kms, and we note that the distinct bipolar pattern provides additional evidence
for the presence a significant component on the plane of the sky. 
Thus, \mysource is very likely the fastest known water fountain
jet.

Proper motion studies will also be key to study the origin of the
remarkably high internal velocity dispersion within each lobe, for
instance, whether it can be explained with projection effects alone.

\acknowledgments

JFG and OS  acknowledge partial support from Ministerio de Ciencia e
Innovaci\'on (MICINN) of Spain, grant AYA2008-06189-C03-01. OS, LFM, MAG, and GR-L are
partially funded by MICINN grant AYA2008-01934. JRR acknowledges support by MICINN grants AYA2009-07304 and CSD2009-00038.
JFG, LFM, and MAG are also
supported from Junta de Andaluc\'{\i}a (TIC-126).  GR-L acknowledges
support from CONACyT and PROMEP (Mexico). 
This research is partly based on observations made with the 
Spitzer Space Telescope and the  obtained from the NASA/ IPAC Infrared Science
Archive, both of which are operated by the Jet Propulsion Laboratory,
California Institute of Technology under a contract with NASA, on
observations with AKARI,  
a JAXA project with the  participation of ESA, and on data products
from the Wide-field Infrared Survey Explorer, a joint project
of the University of California, Los Angeles, and the Jet Propulsion
Laboratory/California Institute of Technology, funded by NASA.

{\it Facilities:} \facility{EVLA}, 
\facility{Akari}, \facility{CAO:2.2m}, \facility{IRAS}, 
\facility{MSX}, \facility{Spitzer}

\clearpage

\end{document}